# Can contact potentials reliably predict stability of proteins?

Jainab Khatun[1], Sagar D. Khare[1] & Nikolay V. Dokholyan[*]
Department of Biochemistry & Biophysics, University of North Carolina at Chapel Hill, NC 27599 USA

## Abstract

The simplest approximation of interaction potential between amino-acids in proteins is the *contact potential*, which defines the effective free energy of a protein conformation by a set of amino acid contacts formed in this conformation. Finding a contact potential capable of predicting free energies of protein states across a variety of protein families will aid protein folding and engineering *in silico* on a computationally tractable time-scale. We test the ability of contact potentials to accurately and *transferably* (across various protein families) predict stability changes of proteins upon mutations. We develop a new methodology to determine the contact potentials in proteins from experimental measurements of changes in protein's thermodynamic stabilities ($\Delta\Delta G$) upon mutations. We apply our methodology to derive sets of contact interaction parameters for a hierarchy of interaction models including solvation and multi-body contact parameters. We test how well our models reproduce experimental measurements by statistical tests. We evaluate the maximum accuracy of predictions obtained by using contact potentials and the correlation between parameters derived from different data-sets of experimental $\Delta\Delta G$ values. We argue that it is impossible to reach experimental accuracy and derive fully transferable contact parameters using the contact models of potentials. However, contact parameters can yield reliable predictions of $\Delta\Delta G$ for datasets of mutations confined to specific amino-acid positions in the sequence of a single protein.

---

[1] JK and SDK contributed equally to this work.
[*] Address for correspondences: Email:dokh@email.unc.edu; Phone: (919) 843-2513

# Introduction

A fundamental goal of molecular biophysics is to understand the relationship between protein sequence and structure, a problem known as the *protein folding problem* (1-8). Conversely, identifying amino acid sequences that fold rapidly and are stable in a target conformation, is known as the *protein design problem* (9-20). Despite recent remarkable successes in protein folding *in silico* (21-24), the folding time scales of most proteins are not accessible to detailed computational techniques, such as molecular mechanics simulations. This limitation makes identification of the folding properties of naturally occurring and designed proteins intractable. An alternative approach to the solution of both protein folding and design problems is the development of a simplified model of amino acid interactions that can be used in rapid computational techniques, such as Monte-Carlo simulations (25-27) and discrete molecular dynamics (28-31). Such models incorporate a coarse-grained description of the polypeptide chain where each residue is represented by a limited (one or a few) number of beads connected by a chain. Effective interactions are assigned between beads. The model drastically reduces the computational cost of folding, and is amenable to detailed statistical mechanical characterization.

One of the widely used form of simplified inter-residue potentials is the contact potential in which amino acids interact if they are spatially located within a certain distance from each other (31-38). The *quasichemical* approximation (37) is a method for deriving pairwise contact potentials from the number of residue-residue contacts found in a large set of protein structures. The quasichemical approximation has proved to be successful in designing and folding proteins on lattices (38). A strategy proposed in Ref. (32) to derive pairwise contact potentials for off-lattice protein models was to iteratively refine the contact potentials by starting from a trial set of interaction parameters, and modifying these parameters until the energy minimum conformation corresponded to the native-state of a given protein. Attention has also been paid to find contact interaction parameters such that the energies of the native states of a set of proteins are simultaneously smaller than their respective decoy conformations (33,36). Potentials derived from these methods have successfully captured the main features of the statistical mechanics of proteins but have failed to consistently reach protein native states in folding simulations. The motivation for these studies was to find a contact potential that does not require prerequisite knowledge of protein structures and is capable of folding a model protein to native conformation. On the other hand, the structure-based Gō potential (31,34) has been successful in reaching protein native state in molecular dynamics simulations (28,39). However, since Gō potential is derived in the context of a specific protein native state, it is not transferable to other proteins. A fully transferable universal potential, therefore, remains elusive.

The observed non-transferability of contact potentials raises the questions of whether it is possible to derive a set of contact interactions that are able to (a) discriminate the native states of a set of proteins from all structural decoys, and (b) capture protein thermodynamics to predict the changes in free energy of unfolding of a set of proteins upon single or multi-point mutations with the experimental accuracy? It was shown by Vendruscolo *et al*. (40,41) that no pairwise contact potential exists which can stabilize an off-lattice model of some proteins relative to their structural decoys. The ability of contact models to predict stability changes upon mutations (ΔΔG) has not yet been similarly evaluated. Many successful models for predicting stability changes upon mutations either use detailed representation of the polypeptide chain with continuous potentials (42,43) and/or database-derived statistical potentials (Ref. (44) and references therein).

One approach for deriving interaction parameters for predicting stability changes in proteins upon mutation is to "learn" these parameters by fitting the interaction model to experimental data (43,45). There are two requirements for a derived interaction potential for predicting stability changes of proteins upon mutations: such interaction potential predicts ΔΔG (within experimental error) for (i) the set of proteins from which it is derived (*the criterion of accuracy*), and (ii) sets of proteins that were not used for derivation of this interaction potential (*the criterion of transferability*). Here, we ask whether simple contact models can be used to predict ΔΔG and evaluate the reliability of contact potentials using the criteria of accuracy and transferability. The contact potentials are parameterized from experimental ΔΔG values using the Singular Value Decomposition (SVD) algorithm. We show that contact potentials fail on both the criteria i.e. they are neither accurate nor transferable. We argue that it is impossible to determine accurate transferable contact parameters to accurately predict the changes in stability of a set of proteins simultaneously.

# Methods
## Potential models

We use a hierarchical approach to develop a simplified model of amino acid interactions. We approximate the potential energy of amino acids interactions in a protein by the sum of terms:

$$H = H_2 + H_3 + H_s \ldots, \qquad (1)$$

where $H_2$ and $H_3$ are the contributions from two- and three-body interactions respectively, $H_s$ is the contribution to the protein potential energy from the solvation of amino acids.

Two-body Hamiltonian

We use an interaction model for two-body interaction $H = H_2$, assuming effective interactions between pairs of C$_\beta$ atoms. The two-body Hamiltonian $H_2(\Gamma, S)$ for a protein of conformation $\Gamma$ with a given sequence of amino acids $S$ is the sum of the pairwise interaction potentials $u_2(\sigma_a, \sigma_b)$, between amino acids of types $\sigma_a$ and $\sigma_b$:

$$H_2(\Gamma, S) = \sum_{\sigma_a, \sigma_b} \sum_{i \neq j = 1}^{M, N} u_2(\sigma_a, \sigma_b) s_i(\sigma_a) s_j(\sigma_b) \Delta_{ij}, \qquad (2)$$

where $s_i(\sigma_a) \equiv \delta(\sigma_a - \hat{\sigma}_i)$ and $\hat{\sigma}_i$ is the amino acid type at position $i$ along the sequence $S = \{\hat{\sigma}_1, \hat{\sigma}_2, \ldots, \hat{\sigma}_N\}$; $\delta(x)$ is 1 if x=0 or 0 otherwise; $\|\Delta_{ij}\|$ is the contact matrix whose elements $\Delta_{ij}$ are either 1 or 0 depending on whether amino acids at the positions $i$ and $j$ along the sequence are in contact or not. The contact between amino acids $i$ and $j$ is defined to exist if their corresponding $C_\beta$ atoms ($C_\alpha$ in case of Gly) are separated by a distance of 7.5 Å (38). $M$ is the total number of distinct amino acid types and $N$ is the length of the protein. For the two-body term we account for all $M = 20$ types of amino acids.

Solvation Hamiltonian

Proteins are strongly affected by the solvent, so it is important to take into account the contribution of the solvation energy term, $H_s$ to the total amino acid interactions (46). We use an approximate method for the determination of the solvent contribution originally proposed by Eisenberg (47) which assumes that the effect of solvent is proportional to the solvent-accessible surface area buried upon folding. We approximate the solvent contribution as:

$$H_s = \sum_{\sigma_a}^{M} \sum_{i}^{N} u_s(\sigma_a) s_i(\sigma_a) \Sigma_i , \qquad (3)$$

where $u_s(\sigma_a)$ is the solvent energy term for the amino acid of type $\sigma_a$, and $\Sigma_i$ is solvent accessible surface area for the $i$th amino acid. To compute $H_s$ we calculate the solvent accessible surface area for each amino acid by estimating the surface area swept by the center of a solvent probe molecule (modeled as a rigid sphere of finite radius) when it rolls about the van der Waals surface of the protein. The van der Waals surface is the external surface of the atoms, each represented by a spherical ball of its van der Waals radius. The methodology to perform solvent accessible surface area computations is well-established (47-50). We use freely-available program GETAREA1.1 (48) to estimate the solvent accessible surface area of a residue.

Three-body Hamiltonian

It has been shown that non-additive protein interactions play an important role (51,52) in protein folding. We account for the non-additive interactions by adding terms corresponding to three-body and higher order terms. Using the same notation as in Eq.(2), we approximate the three-body interaction of amino acids as:

$$H_3 = \sum_{\sigma_a,\sigma_b,\sigma_c}^{M} \sum_{i \neq j \neq k=1}^{N} u_3(\sigma_a,\sigma_b,\sigma_c) s_i(\sigma_a) s_j(\sigma_b) s_k(\sigma_c) \Delta_{ij} \Delta_{jk} \Delta_{ki} , \qquad (4)$$

where $u_3(\sigma_a, \sigma_b, \sigma_c)$ are the interaction parameters between amino acids of types $\sigma_a$, $\sigma_b$, and $\sigma_c$. Eq.(4) requires the formation of a triad of contacts simultaneously between amino acids at positions $i$, $j$, and $k$, so that $H_3$ contributes to $H$ only when there are triads of amino acids that are in contact with each other. The $H_3$ term has, therefore, an additional contribution favoring or disfavoring triad formations. The $H_3$ and higher-order non-additive terms drastically increase the number of interaction parameters: for $M = 20$, the number of interaction parameters in $H_2$ (Eq.(2)) is 210, while the number of parameters in $H_3$ (Eq.(4)) is 1540. We also use a reduced alphabet of amino acids (53) $M = 6$ in $H_3$. In the reduced alphabet, we group amino acids (53,54) to aliphatic (A, V, L, I, M and C), aromatic (F, W, Y and H), polar (S, T, N and Q), positively charged (K and R), negatively charged (D and E), and special (reflecting their special conformational properties) (G and P). We have a total of 56 reduced-alphabet three-body interaction parameters.

## Deriving the interaction parameters

To find the interaction parameters ($u_2(\sigma_a, \sigma_b)$, $u_3(\sigma_a, \sigma_b, \sigma_c)$, $u_s(\sigma_a)$,...) in Eqs. (1) – (4) we use experimentally determined changes in the thermodynamic stabilities of proteins upon mutations ($\Delta\Delta G$) from a large number of mutant proteins. Our derivation is based on the approximation that the changes in entropies between mutant and wild type proteins predominantly arise due to the differences in the solvation free energies of the mutant and wild type amino acids, so that the structures of both mutant and wild type proteins remain close to each other (within 2 Å RMSD), enough to assume that the change in conformational entropy is zero (we note that the solvation free energy difference can be accounted by the term $H_s$ (Eq.(3)). Thus, the difference between stabilities of mutant and wild-type proteins is due to the change in contact energies arising due to mutation. As contact energies are effective free energies, from the quasi-chemical approximation, we can equate the experimentally determined changes in thermodynamic stabilities $\Delta\Delta G$ to $\Delta\Delta H$ values from our model, i.e. $\Delta\Delta G \approx \Delta\Delta H = \Delta H_{mut} - \Delta H_{wt} = H_{mut} - H^u_{mut} - H_{wt} + H^u_{wt}$, where $H_{mut}$, $H^u_{mut}$, $H_{wt}$, and $H^u_{wt}$ are the energies of mutant native and unfolded and wild type native and wild type unfolded states correspondingly. We consider a model of the unfolded state in which mutation does not affect the intra-protein contacts in the unfolded state. Since the difference in energies of the unfolded states can be approximated by the changes in the solvation free energies of the wild-type and mutant amino-acids, accounted by Eq.(3), we neglect the difference $H^u_{mut} - H^u_{wt}$ and obtain

$$\Delta\Delta G = H_{mut} - H_{wt}. \qquad (5)$$

By substituting Eqs.(1)-(4) to Eq.(5), for each $\Delta\Delta G$ measurement, we obtain a linear equation relating experimentally determined free energy change and contact energy parameters of our model with a given alphabet size (e.g. 210 two-body, 1540 three-body and 20 solvation interaction parameters for a 20-letter amino acid alphabet). For all the measurements in a particular dataset, we obtain a set of linear equations relating the free energy changes and the contact interaction parameters. We simultaneously solve the set of linear equations using Singular Value Decomposition (SVD) corresponding to a large number of mutants to determine these interaction parameters.

## Singular Value Decomposition

Since the number of $\Delta\Delta G$ measurements may not be exactly equal to the total number of variables (the interaction parameters $u_2(\sigma_a, \sigma_b)$, $u_3(\sigma_a, \sigma_b, \sigma_c)$, $u_s(\sigma_a)$,... in (Eqs. (1)- (4)), it is expected that *exact* solutions are not achievable. Instead, we find the *linear least square* solution of Eqs (5). The *linear least square problem* for a set of linear equations, $A\vec{x} = \vec{b}$, where $A$ is a known $m \times n$ matrix, $\vec{b}$ is a known $m$-dimensional vector and $\vec{x}$ is a $n$-dimensional solution vector, is defined as finding a solution vector $\vec{x}$ that minimizes the L$_2$ norm of the residual vector $\| A\vec{x} - \vec{b} \|_2$. If $m \geq n$ and $rank(A) = n$, we find a unique least square solution. If $m < n$, there is an infinite number of least square solutions, so we seek the minimum norm least square solution which minimizes both $\| \vec{x} \|_2$ and $\| A\vec{x} - \vec{b} \|_2$.

To find the solution for both the above cases, we implement the SVD algorithm (55). The SVD of a $m \times n$ matrix $A$ is given by $A = UWV^T$, where $U$ and $V$ are orthogonal matrices and $W$ is a diagonal matrix whose elements $w_i$, known as singular values of $A$, satisfy $w_1 \geq w_2 \geq \ldots \geq w_{min} \geq 0$. The SVD algorithm ensures a unique decomposition of this matrix. In our case, $A$ corresponds to the matrix of contact interactions, $\vec{x}$ is the solution vector of contact energies and $\vec{b}$ is the experimentally determined $\Delta\Delta G$ values. The dimension of $\vec{x}$ and $\vec{b}$ are the number of parameters ($N_p$) to be determined and the number of experimental measurements ($N_e$) respectively. We identify an approximate solution that satisfies Eqs.(5) for all the mutants simultaneously so that

$$\chi^2 \equiv \sum_{\alpha=1}^{N_e} \| \Delta\Delta G - \Delta\Delta H \|_\alpha^2, \qquad (6)$$

is minimized ($\chi^2_{min} = \min \chi^2$). In Eq.(6) $\alpha$ enumerates $N_e$ experimental measurements.

In the SVD algorithm, we decompose the matrix $A$ into its singular values, $\{w_i\}$. If the matrix is singular we find that some $w_i$ are zero. If the matrix $A$ is not singular, but ill-conditioned (i.e. some $w_i$ values cannot be accurately determined), then the solution vector may have a large error-prone component which yields a non-least square solution ($\chi^2 \neq \chi^2_{min}$). Therefore, in order to find the least square solution for this case we define the *relative singular value*, $\varepsilon_i = w_i / w_{max}$ ($w_{max}$ is the maximum singular value in magnitude). If $\varepsilon_i$ is smaller than some *threshold* value, $\varepsilon$, we neglect the corresponding $w_i$ for finding the solution. Thus, $\varepsilon$ determines the number of non-zero singular values in (and consequently the rank of) the design matrix $A$ and hence the solution vector $\vec{x}$. In order to obtain the most accurate solution, we choose the value of $\varepsilon$ such that the value of $\chi^2$ is the minimal ($\chi^2_{min}$). If multiple solutions correspond to a value of $\chi^2_{min}$, we choose the minimal norm (minimizing the length of $\vec{x}$) solution.

## Tests of reliability of models
Resubstitution test

The *resubstitution test* is an examination for the self-consistency of a prediction algorithm. In the resubstitution test *ΔΔG* for each mutant are predicted for the dataset (*testing set*) from which the model parameters are derived (*training set*). The resubstitution test sets an upper bound on the prediction ability of the derived potential when the training and testing sets are not identical. The performance of a model in the resubstitution test is an optimistic estimation (56-58). Thus, the resubstitution test is necessary but not sufficient for evaluating a prediction method. As a complement, a cross validation test is needed because it would reflect the effectiveness of a prediction algorithm.

Jack-knife test

We perform jack-knife test, also known as leave-one-out test (59), to cross-validate in statistical prediction of our interaction model. The jack-knife provides an objective assessments (60) of the predictive capabilities of a model. In the jack-knife test the experimental data of *N* points is divided into two groups: the training set consisting of *N*-1 data points and the testing set consisting of 1 data point. We derive the interaction model parameters using the training set data and predict the value of $\Delta\Delta G$ for the testing set. Then another data point is selected from the experimental data to represent the testing set, while the rest *N*-1 data points represent training set. We repeat this procedure for all *N* data points. The training and testing, thus covers all of the experimental data points.

## Results and Discussion

We derive a hierarchy of amino acid interaction potentials from several datasets of experimentally measured changes in free energy upon mutation (Methods). In order to test accuracy as well as transferability of potentials, we choose three different datasets (Table I). (i) DS1 consists of 303 four-point mutants of eglin c, a small 70-residue, protease inhibitor structurally homologous to chymotrypsin inhibitor 2. The $\Delta\Delta G$ values for eglin c were determined with high accuracy (upto ±0.087 kcal/mol) by Edgell *et al.* (61) for (quadruple) mutations at R22, E23, T26 and L27 in the solvent-exposed helix of eglin c. (ii) DS2 consists of 658 single, double, triple and quadruple mutants for staphylococcal nuclease (149 amino acids). We use $\Delta\Delta G$ values reported for staphylococcal nuclease in the ProTherm database (62) and in Zhou *et al* (63). The typical error bar reported for staphylococcal nuclease mutant $\Delta\Delta G$ values are approximately ±0.1 kcal/mol. The mutation sites are not localized to any particular sub-structure of the protein. (iii) DS3 consists of 1356 $\Delta\Delta G$ measurements for a set of eleven proteins that are reported in ProTherm (62), Guerois *et al.* (43) and Zhou *et al.* (63). The mutation sites are scattered throughout the structure of these eleven proteins. The maximum error bar for this dataset is 0.48 kcal/mol. We derive potentials for each data set and compute the accuracy of our predictions by evaluating the models' predictive capability. We also test the ability of potentials derived from one dataset to predict changes in stability of other datasets (transferability).

## Predicting $\Delta\Delta G$ using contact models
Two-body interaction potential

We consider pairwise two body interaction ($H_2$, Eq.(2)) and derive interaction parameters $u_2(\sigma_a, \sigma_b)$ (210 parameters for a 20-letter amino-acid alphabet) from a set of equations (such as Eq.(5)) corresponding to all the mutants in a given dataset, by employing the SVD algorithm. Using derived interaction parameters, we predict the change in free energy ($\Delta\Delta G$) upon mutation and examine our prediction quality by both the resubstitution and the jack-knife tests (see Methods) for all the three datasets.

The resubstitution test is crucial because it reflects the self-consistency of the prediction algorithm and sets an upper bound on the performance of a given model. When the experimental dataset is limited, the resubstitution test is useful for assessment of the accuracy of an interaction model (64), while the jack-knife test provides a rigorous assessment of the model's predictive power. In both tests, we compute the linear regression correlation coefficients between the predicted and actual experimental values (Table II). For the more accurate eglin c measurements (DS1), we obtain resubstitution ($r_s$) and jack-knife ($r_j$) correlation coefficients of 0.87 and 0.80 respectively (Figure 1(a)), while the values for the same quantities are 0.79 and 0.52 for DS2 (Figure 1(b)) and 0.66 and 0.46 for DS3 (Figure 1(c)).

The correlation between the predicted and actual values of $\Delta\Delta G$ is expected to decrease in the jack-knife test in comparison with the resubstitution test. For both tests, we observe that there are common outliers for which predicted $\Delta\Delta G$ values are significantly different than the experimental ones. Specifically, for DS1, the outliers are the four four-point mutants (R22K/E23K/T26N/L27K, R22N/E23K/T26N/L27A, R22E/E23G/T26E/L27G and R22P/E23H/T26E/L27K) for which the $\Delta\Delta G$ values are 2.488, 2.623, -5.676 and -5.698. Most experimentally measured $\Delta\Delta G$ values in DS1 lie between approximately -3.0 and 0.5 kcal/mol (the average and the standard deviations of $\Delta\Delta G$ measurements are -1.393 and 0.83 respectively). Such large perturbations of free energy in the outliers indicate large structural alterations which, in turn, imply that the contact approximation is not valid in these cases. If the outliers are ignored, the values of $r_s$ and $r_j$ are 0.93 and 0.89, respectively.

The significantly better performance of the model for DS1 compared to DS2 and DS3 can be partially attributed to greater accuracy (error of measurement = 0.087 kcal/mol) and greater consistency in experimental measurements (all data points are obtained from one source and at the same experimental conditions). The stability determination of any given protein is a function of the conditions under which the measurements are made. For example, pH and salt concentrations can alter the measured stability of a particular protein dramatically (65-67).

Apart from the variability of sources of error in experimental data, the statistical properties of the dataset also determine the performance of a model in the jack-knife test. The more a given interaction parameter is represented in the set of linear equations, the less likely is its estimation to significantly change in the jack-knife test. We estimate the representation of parameters in the dataset using the *coverage parameter*, $C_p$, which we define as:

$$C_p = \frac{1}{N_p} \sum_{f=0}^{f=1} f n_f \qquad (7)$$

where $n_f$ is the number of parameters occurring with frequency $f$ and $N_p$ is the total number of parameters. For a dataset with all parameters represented in all equations, $C_p=1$. Small values of $C_p$ indicate that a large number of parameters are under-represented in the set of equations corresponding to the dataset. The $C_p$ values for datasets DS1, DS2 and DS3 are 0.349, 0.045 and 0.044, respectively (Fig.2). Thus the jack-knife performance is directly related to the parameter space coverage.

In addition, we note that DS1 consists exclusively of mutations localized to four specific positions on eglin c (R22, E23, T26 and L27). DS2 consists of mutations in one protein staphylococcal nuclease, albeit scattered throughout its structure; DS3 consists of a variety of mutations in a variety of proteins. Thus, DS2 and DS3 contact parameters average over diverse structural environments. We find that contact energies derived from mutations that are localized within similar structural environment (e.g. mutants on the solvent-exposed helix in eglin c, DS1) are more successful in self-consistently predicting the respective energetics than contact energies derived from averaging over a large set of environments and conditions (as in DS2 and DS3). The above observation confirms that contact energies are not independent of mutants' structural environments suggesting that including additional terms in the interaction model may capture sufficient environmental detail to successfully predict protein stability.

Adding solvent and multi-body terms

It has been argued that solvation and multi-body effects are important to capture structural environments of amino acids (46,52,68). Therefore, we add the solvation effect Eq.(3) in our potential model and derive the interaction with solvent $u_s(\sigma_a)$ along with two-body interaction parameters $u_2(\sigma_a, \sigma_b)$. Following Ref. (47), we estimate the solvation energy as a linear function of the solvent accessible surface area of the wild-type residue. There are a total of 230 (20 solvation + 210 two-body) parameters for the 20-letter amino acid alphabet size. We predict the change in free energy upon mutation and examine the prediction quality of the new derived potential (Table II). We obtain correlation coefficient for self-consistency and jack-knife tests to be 0.80 and 0.78 for the dataset DS1, 0.80 and 0.47 for DS2 and 0.65 and 0.45 for DS3 respectively. We find that for all databases, the inclusion of the solvation term does not significantly affect either the self-consistency or jack-knife tests performance. An accurate treatment of solvation considering atomistic details of the polypeptide chain may increase the possibility of finding better correlation between experimental and predicted values of free energy changes.

Finally, we add the three body term $H_3$ (Eq. (4)) in our interaction potential model, so that the resulting Hamiltonian consists of the terms $H_2$ (Eq. (2)), $H_s$ (Eq. (3)) and $H_3$ (Eq. (4)). There are a total of 1770 parameters (210 two-body, 20 solvation and 1540 three-body) are to be determined for a twenty letter alphabet size. The correlation coefficient including three-body interaction for self-consistency and jack-knife tests are 0.87 and 0.78 for DS1 0.98 and 0.28 for DS2 and 0.91 0.11 for DS3 respectively. We find that the three-body Hamiltonian performs better for the self-consistency test but leads to a worse performance on jack-knife tests. This observation can be rationalized as follows: we need to determine 1770 parameters on inclusion of three-body interactions while we have 303, 658 and 1356 equations in DS1, DS2 and DS3 respectively. Therefore, the set of equations corresponding to $H_3$ represents an underdetermined set with more variables than the number of equations. The parameter space coverage for the $H_2+H_s+H_3$ Hamiltonian does not significantly change from when we consider only $H_2$ Hamiltonian: the $C_p$ values are 0.34, 0.02 and 0.01 respectively for DS1, DS2 and DS3 (Figure 3).

It is possible that the performance of $H_2+H_s+H_3$ Hamiltonian model is a reflection of the limited number of data points in any dataset used to calculate the energy parameters; consequently, these experimental datasets are not large enough for predicting all 1770 parameters. Sufficient sampling of the parameter space can be obtained by rationally designed large-scale mutant libraries. While the inclusion of multi-body effects may increase the self-consistency and jack-knife tests correlations, we argue below that the solutions may not be universal, i.e. we will not be able to predict stability changes accurately on datasets other than the one from which the parameters are derived.

In order to circumvent the paucity of accurate data, we use a reduced amino acid alphabet to obtain a smaller number of parameters. The minimal amino acid alphabet for protein folding has been an active area of research (53,69,70). Recent studies have suggested (53) that the full sequence complexity is not required to design a foldable protein and an alphabet size of 6 may be sufficient to fold a protein. Thus, we reduce the number of parameters using six types of amino acids (Methods) for the three body interaction; the two-body and the solvent effects are parameterized for the twenty amino acid alphabet. There are a total of 286 parameters to be determined (210 two-body and 20 solvation terms for 20 letter alphabet, 56 three-body terms for 6-letter alphabet size). The performance of this partially reduced alphabet model is not significantly different from the previous model with the full 20-letter amino acid alphabet. The histograms (not shown) of parameter space for this model indicate that the reduction of the alphabet size does not lead to significant improvement in the parameter coverage due to the mutual cancellation of parameter terms in the equations.

## Accuracy of parameters derived from contact models

The relatively high correlation coefficients obtained, ($r_s$=0.87; with $H_2$ for DS1), indicate that a high proportion data can be explained by our model. However, the measure of a model's predictive capability is the uncertainty in the predictions compared to the experimental errors. Therefore, we analyze how accuracy of $\Delta\Delta G$ predictions and compare this accuracy with the experimental uncertainty.

As described in Methods, we solve a set of linear equations $A\vec{x} - \vec{b} = 0$, where $A$ corresponds to the matrix of contact interactions, $\vec{x}$ is the solution vector of contact energies and $\vec{b}$ is the experimentally determined $\Delta\Delta G$ values. Using SVD algorithm, we find the linear *least square solution*, the solution vector for which $\chi^2 = \|(A\vec{x} - \vec{b})\|_2 / N_e$ is minimal ($\chi^2_{\min}$), where $N_e$ is the number of experimental measurements. $\chi^2$ depends on a tunable threshold parameter ε, which determines the number of non-zero singular eigen-values of the matrix $A$. For ε in the range $10^{-3}$-$10^{-14}$, the $\chi^2$ is constant and minimal (see Fig. 4). Therefore, we use a value of ε corresponding to the above range (ε=$10^{-8}$). With this value of ε we obtain the solution vector and the associated $\chi^2_{\min}$. The value of $\chi^2_{\min}$ is the parameter that represents the average uncertainty in our prediction, and is therefore the relevant parameter to compare with experimental uncertainties in measurement.

For all the datasets, we find that the uncertainty in our predictions ($\chi^2_{\min}$) is greater than the observed experimental error. Thus, we can never reach $\chi^2_{\min} \leq \chi^2_{\exp}$, where $\chi^2_{\exp}$ is the square of the experimental error-bar. The $\chi^2_{\min}$ for dataset DS1 (using only two-body Hamiltonian) is approximately 0.40 kcal/mol, while the experimental error bar is 0.087 kcal/mol (61,71) (Fig 4(a)). For the dataset DS2 we obtain $\chi^2_{\min}$ to be equal to 0.9 kcal/mol (Fig. 4(b)), while the experimental error bar is approximately 0.1 kcal/mol (65,72,73). For the dataset DS3 we obtain $\chi^2_{\min}$ to be equal to 1.31 kcal/mol (Fig. 4(b)), while the maximal experimental error bar is approximately 0.48 kcal/mol (74) (Fig 4(c)). We obtain similar results upon inclusion of the other interaction terms ($H_s$, $H_3$) (data not shown). SVD, with an appropriate choice of the ε parameter, is *guaranteed* to yield the most accurate *least square solution* for a given dataset. Thus, for all datasets, we show that the uncertainties in prediction are always greater than the corresponding experimental error.

## Transferability of parameters derived from contact models

In order to derive a universal set of contact interaction parameters, it is crucial that parameters derived from one dataset accurately predict the $\Delta\Delta G$ values for other datasets. Therefore, we analyze the transferability of interaction parameters between the datasets. We determine the correlation coefficients between the parameters derived from different sets using the two-body Hamiltonian and find that the correlation is negligible (Fig.5). The lack of correlation between parameters that include solvent and the multi-body effects derived from different datasets (data not shown) also demonstrates the non-transferability of interaction parameters.

The uniqueness of the derived solution depends on the relative number of parameters ($N_p$) and number of experimental observations ($N_e$). There are three possibilities to consider for evaluating the uniqueness of a set of contact parameters.

<u>Case 1:</u> $N_e < N_p$. This case represents an under-determined set of equations, so that we always find more than one least square solution. Among these different solutions we choose the minimal norm solution vector corresponding to $\chi^2_{\min}$.

<u>Case 2:</u> $N_e > N_p$. This case represents an over-determined set of equations so that no exact solution exists. We find the *least square solution* which is the closest to satisfying all the equations simultaneously (by minimizing $\chi^2$). However, we find that the number of non-zero singular eigenvalues and, hence, the rank of the matrix A ($N_k$) is less than the number of parameters ($N_p$) to be determined. Thus, there is more than one set of solution vectors x that correspond to the value of $\chi^2_{\min}$. The number of least square solution vectors is determined by the eigenvalue threshold ε (Fig. 6). When ε=0, which corresponds to considering *all* the singular values in the SVD procedure ($N_k = N_p$), we obtain a unique solution. As we increase the value of ε, $N_k$ becomes smaller than $N_p$, and we obtain a number of solutions. If each choice of $N_k$ parameters for a given value of $N_p$ corresponds to a different solution compatible with a given value of $\chi^2_{\min}$, then we estimate the number of least square solutions as the number of ways of choosing $N_k$ parameters out of $N_p$. When ε=1, $N_k$=1, corresponding to solving one equation, and the number of solutions is $N_p$.

<u>Case 3</u>: $N_e = N_p$. This case results in a unique solution if the set of equations is not redundant. However, for our datasets, the solution depends on which measurements we choose to derive the parameters from, so that we obtain different solutions depending on the choice of mutants. If the equations are redundant, then finding the least square solution reduces to case 2 and there is more than one solution in this case.

For all the above cases 1—3, we obtain more than one least square solution. We summarize the balance between uniqueness and accuracy of the derived solutions in a cartoon representation in Fig. 7. As stated earlier, the threshold value ε determines the number of non-zero singular values and, hence, both the accuracy and the number of solutions. For a specific value of ε ($\varepsilon=\varepsilon_{min}$), the value of $\chi^2$ is minimal. For $\varepsilon < \varepsilon_{min}$, we include smaller singular values in the solution, so that the components of the solution vector corresponding to these singular values are error-prone (See Methods) and yield $\chi^2 > \chi^2_{\min}$. For $\varepsilon > \varepsilon_{min}$, we discard some significant singular eigenvalues, and the number of effective parameters is not sufficient to fit the data with experimental accuracy. As a result, we again obtain $\chi^2 > \chi^2_{min}$. Thus, the accuracy of obtained solutions has a minimum at $\varepsilon_{min}$, and increases for all other values of $\varepsilon$. The uniqueness of solutions is also a function of $\varepsilon$. We obtain a unique solution when $\varepsilon=0$, and $N_p$ solutions when $\varepsilon=1$ (Figs. 6 and 7). However, for both cases, the corresponding values of $\chi^2$ are much larger than $\chi^2_{min}$. For *all* datasets, when $\varepsilon$ is minimal ($\varepsilon=\varepsilon_{min}$), which corresponds to the highest accuracy of prediction, the number of solutions is much greater than one. Therefore, no unique solution exists and we argue that transferability of parameters from one dataset to the other can not be achieved.

## Implications for the assumptions and usefulness of the contact model

The implausibility of deriving an accurate and transferable contact potential offer insights into the assumptions behind the contact models used and their utility. Two assumptions implicit in the contact approximation— first, that contact energies are independent of the protein environment, and, second, that the unfolded states are unstructured and not affected by mutation—may be violated leading to the observed results.

We observe that when contact energies are derived using studies performed under same conditions (e.g. temperature, pH, and salt concentration) for mutations at the same residues in a single protein (DS1 for eglin c), we obtain a greater accuracy of prediction in both the self-consistency and jack-knife tests than for heterogeneous DS2 and DS3 datasets. This implies that contact energies are a function of the environment in which a given contact is formed. Thus, the usefulness of contact models is dependent on the experimental database under consideration. For a dataset of mutants corresponding to identical positions in the sequence of a single protein (dataset DS1), contact potentials can reliably predict the stability changes upon mutation (correlation coefficient of 0.79 in the jack-knife test) and are therefore reasonably adequate for prediction of similar mutants in the protein. However, for datasets comprising of either mutants corresponding to several distinct amino-acid positions in the sequence of a given protein (dataset DS2) or a set of proteins (dataset DS3) the corresponding correlation coefficients in the jack-knife test are 0.52 and 0.46, indicating their predictive inadequacy.

Furthermore, it has been noted (75-78) that there is residual structure in the unfolded states, and it is likely that effect of mutations on unfolded states may significantly contribute to the protein stability. For these proteins, the assumption of attributing the $\Delta\Delta G$ exclusively to the change in contact energies from the folded state breaks down. A quantitative understanding of the energetics of the unfolded states may therefore be necessary for understanding protein stability.

## Conclusions

We develop a new methodology for deriving contact interaction parameters to predict stability changes in proteins upon mutation ($\Delta\Delta G$). Using experimental $\Delta\Delta G$ measurements, we derive contact interaction parameters for energy functions that include two-body, three-body and solvation contributions to protein stability, by solving a set of linear equations. Each linear equation relates the changes in the amino-acid contact composition of the protein upon mutation to the experimentally measured $\Delta\Delta G$ value. We obtain a reasonable correlation between experimental and predicted $\Delta\Delta G$ with a simple two-body contact approximation and find that the inclusion of solvation and multi-body interactions does not lead to a significant improvement in predictive capability of our models. The differences in the predictive capabilities are related to the frequency of occurrence of contact parameters in the set of linear equations being solved. Therefore, rationally designed mutant libraries, intended to maximally cover parameter space, will aid the estimation of contact parameters. However, we argue that even for well-represented datasets, the $\Delta\Delta G$ predictions may not be adequately accurate. For all datasets in our study – corresponding to $\Delta\Delta G$ measurements for individual proteins (eglin c, staphylococcal nuclease) and for a set of eleven proteins – we find that the uncertainties in predictions made by using the contact approximation are always greater than the corresponding experimental error-bars. Furthermore, we compare interaction parameters derived from the three datasets and find negligible correlation between them. We show that the number of optimal solutions which satisfy all three datasets is always greater than one. Thus, we argue that it may not be possible to determine universal, fully transferable contact interaction parameters that will be able to accurately and simultaneously predict $\Delta\Delta G$ for a set of proteins. Our results suggest that a more atomistic form of potential and/or inclusion of the unfolded state into contact models may be necessary for both an accurate estimation of protein stability and achieving transferability of derived potentials across a set of proteins.

## Acknowledgements
We thank M. Edgell for providing eglin c stability data (61) prior to publication and F. Ding, B. Kuhlman, and E. I. Shakhnovich for helpful discussions.

**Table *I***: Summary of datasets that we use to derive the contact interactions in proteins.

| DataSet | Protein/PDB Code | Number of mutants | Source | References |
|---|---|---|---|---|
| DS1 | Eglin c (1EGL) | 303 | Edgell *et al.* | (61,71) |
| DS2 | Staphylococcal nuclease (1STN) | 658 | ProTherm database Guerois *et al.*, Zhou *et al.* | (43,62,63,79) |
| DS3 | 1STN, 1ARR, 1BNI, 1BPI, 1BP2, 1BVC, 1C90, 1FKJ, 2ABD, 2LZM, 1EGL | 1356 | Edgell *et al.* ProTherm database, Guerois *et al.*, Zhou *et al.* | (43,61-63,71,79) |

**Table *II***: The correlation coefficient between predicted and experimental values of $\Delta\Delta G$ for self-consistency ($r_s$) and jack-knife ($r_j$) tests using Hamiltonians of various complexities. $H_2$, $H_3$, and $H_s$ denote two- and three-body and solvation Hamiltonian interaction terms respectively. $p$ is the measure of the statistical significance of the correlation coefficient (the probability of observing these correlations by chance).

| Protein | Number of mutants | Terms included in $H$ | Self-consistency test | | Jack-knife test | |
|---|---|---|---|---|---|---|
| | | | $r_s$ | $\log_{10}p$ | $r_j$ | $\log_{10}p$ |
| eglin c | 303 | $H_2$ | 0.87 | -52 | 0.80 | -44 |
| | | $H_2+H_s$ | 0.87 | -52 | 0.78 | -42 |
| | | $H_2+H_s+H_3$ | 0.87 | -52 | 0.78 | -42 |
| | | $H_2+H_s+H_3{}^*$ | 0.87 | -52 | 0.77 | -41 |
| staphylococcal nuclease | 658 | $H_2$ | 0.79 | -91 | 0.52 | -39 |
| | | $H_2+H_s$ | 0.80 | -92 | 0.47 | -33 |
| | | $H_2+H_s+H_3$ | 0.98 | < -100 | 0.28 | -12 |
| | | $H_2+H_s+H_3{}^*$ | 0.88 | < -100 | 0.47 | -33 |
| combined | 1356 | $H_2$ | 0.66 | < -100 | 0.46 | -64 |
| | | $H_2+H_s$ | 0.65 | < -100 | 0.45 | -62 |
| | | $H_2+H_s+H_3$ | 0.91 | < -100 | 0.11 | -05 |
| | | $H_2+H_s+H_3{}^*$ | 0.69 | < -100 | 0.45 | -62 |

---

\* indicates that the three-body interactions are calculated using only six different types of amino acids

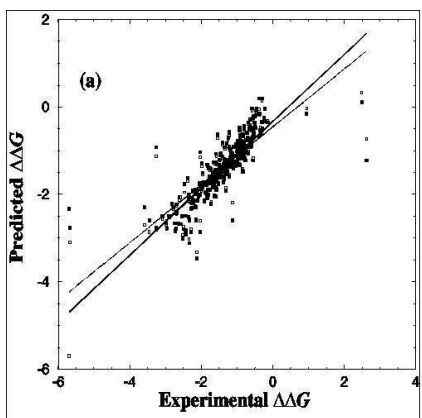
Fig. 1(a)

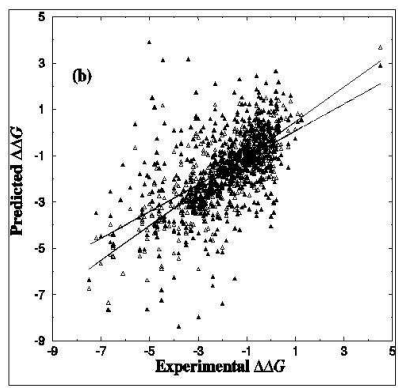
Fig. 1(b)

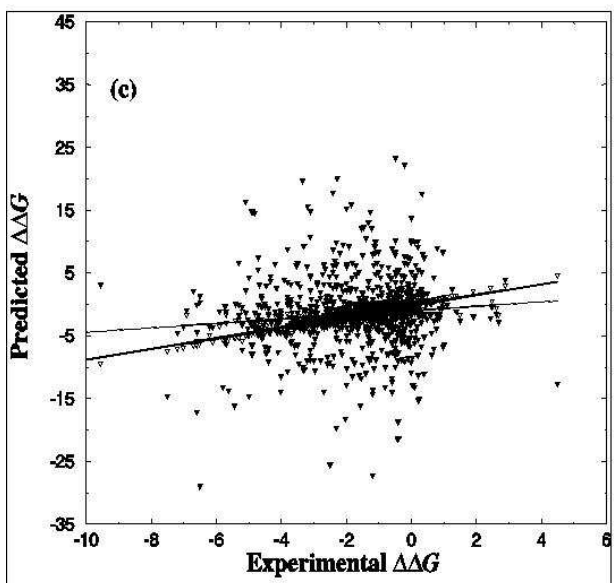
Fig. 1(c)

**Figure 1.** Calculated $\Delta\Delta G$ values compared to the experimental $\Delta\Delta G$ ones: (a) for 303 mutants of eglin c (dataset DS1) for self-consistency (■) and jack-knife (□) tests, (b) for 658 mutants of staphylococcal nuclease (dataset DS2) for self-consistency (▲) and jack-knife (△) tests, and (c) for 1356 mutants of a combined set of eleven proteins (dataset DS3) for self-consistency (▽) and jack-knife (▼) using 2-body Hamiltonian. The two solid lines represent the linear regression obtained for the self-consistency (thick) and the jack-knife (thin) tests for all the datasets. The correlation coefficient ($r$) and the probabilities that this correlation is observed at random ($p$-values) for these tests are reported in Table II.

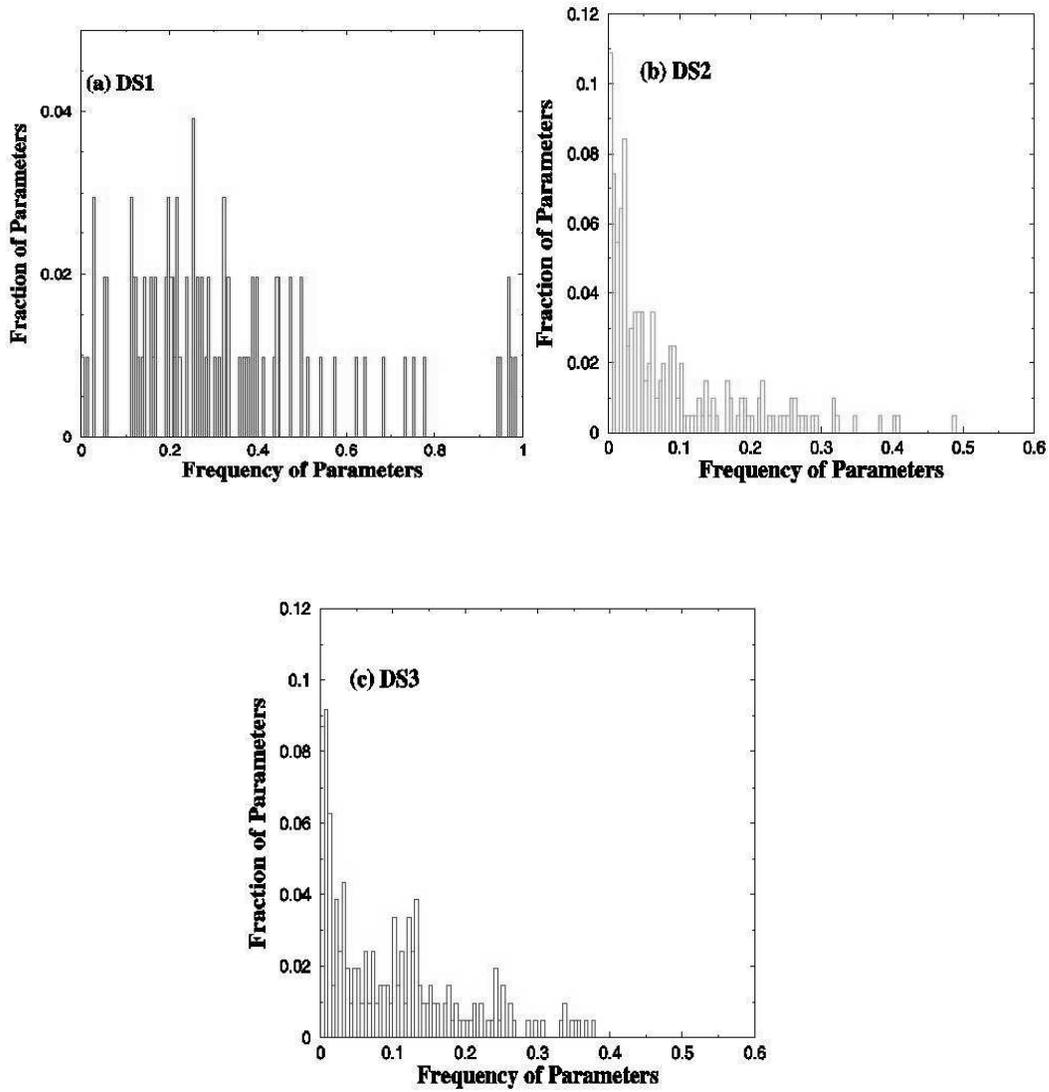

**Figure 2.** The fraction of occurring parameters as a function of frequencies for the two-body interaction Hamiltonian ($H=H_2$): for (a) DS1, (b) DS2 and (c) DS3 respectively.

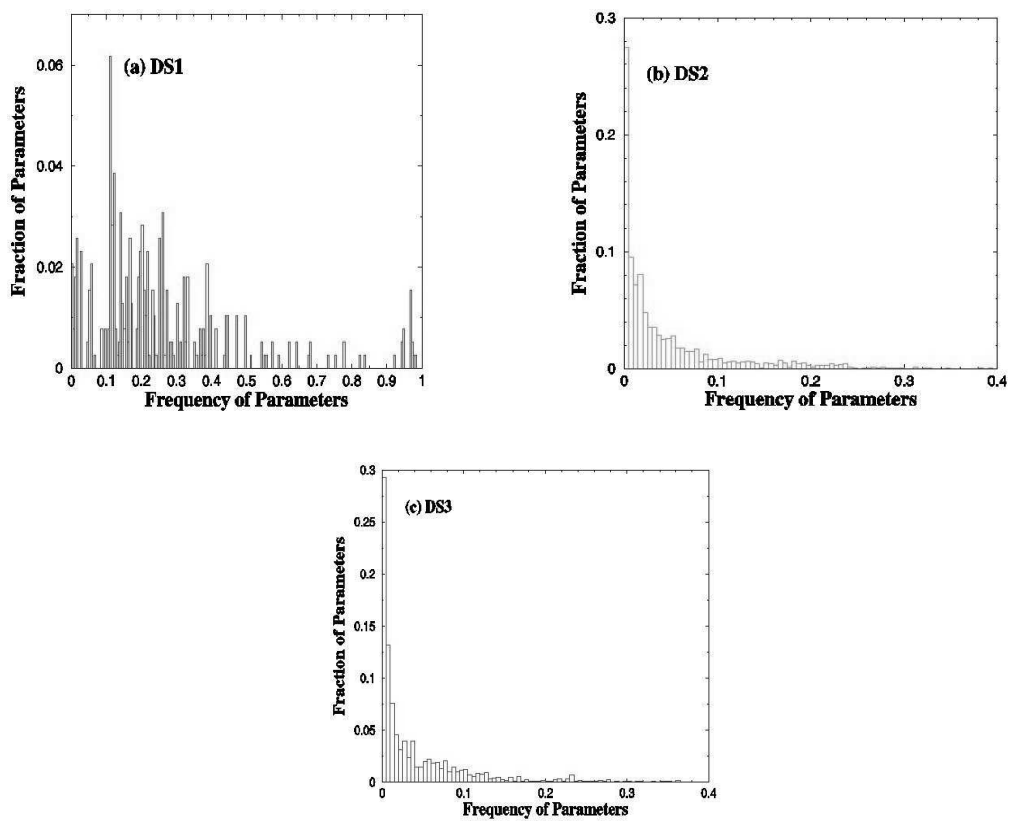

**Figure 3.** The fraction of parameters as a function of their frequency of occurrence for datasets (a) DS1, (b) DS2 and (c) DS3 respectively. The histograms are obtained for interaction models that include two- and three-body and solvent interactions terms ($H= H_2+H_s+H_3$). The comparison of Figs. 2 and 3 indicates that the inclusion of three-body interactions reduces the parameter space coverage in the composite Hamiltonian ($H= H_2+H_s+H_3$) compared to the two-body Hamiltonian ($H= H_2$) parameter coverage.

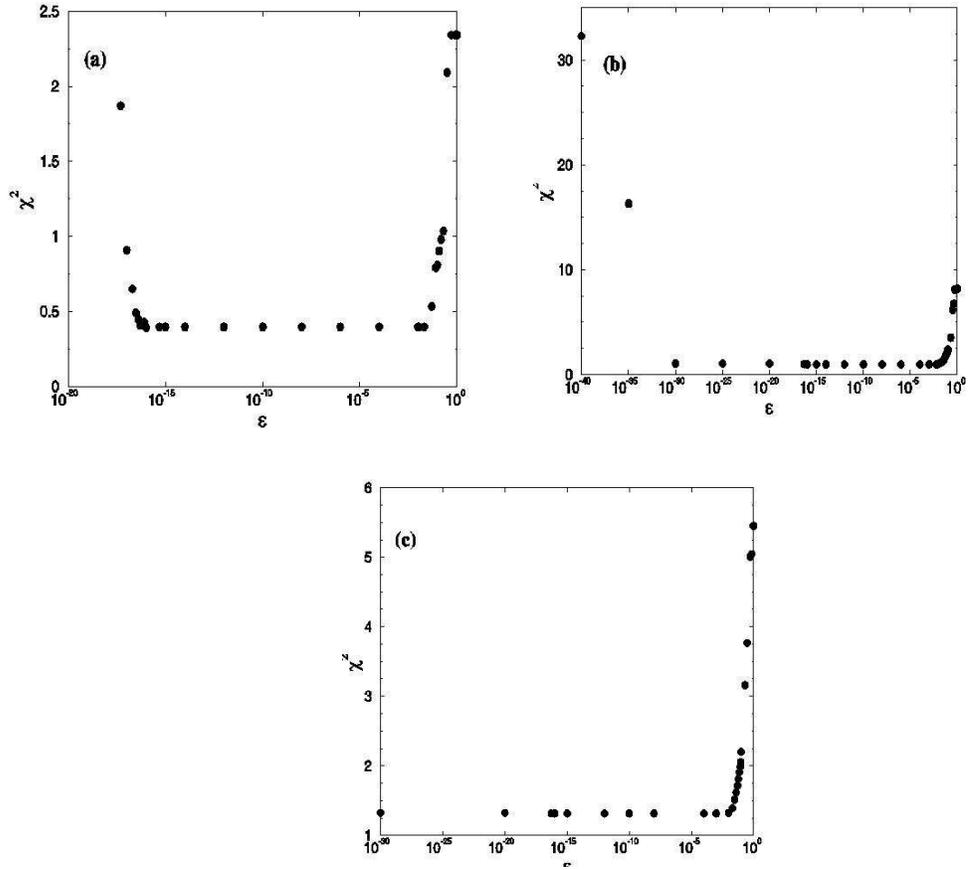

**Figure 4.** The dependence of $\chi^2$ on $\varepsilon$ for datasets (a) DS1, (b) DS2, and (c) DS3. The $\chi^2$ values remain constant for a range of $\varepsilon$ ($10^{-3}$ to $10^{-14}$) and increases for values of $\varepsilon$ outside this range. For, DS3 the increase for smaller values of $\varepsilon$ is less pronounced. For each dataset, we calculate the linear least square solution at $\varepsilon = 10^{-8}$ and obtain the minimal $\chi^2$ value.

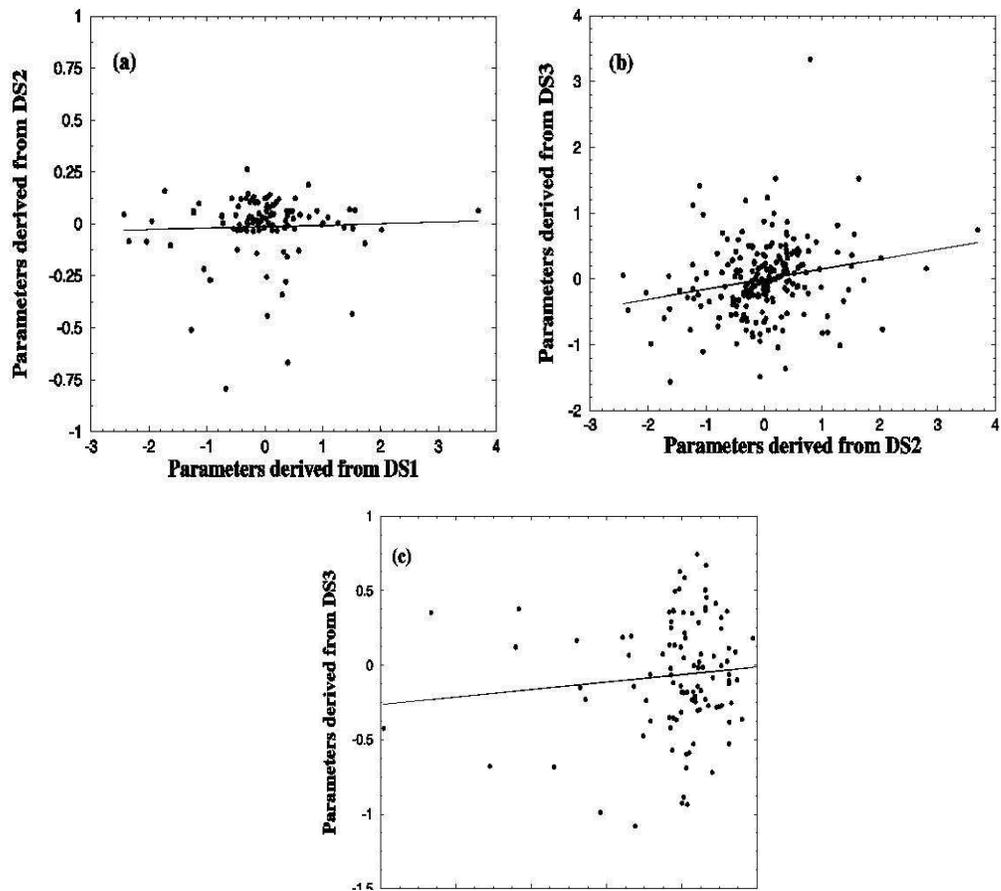

**Figure 5.** The comparison of the interaction parameters derived from datasets (a) DS1 and DS2, (b) DS2 and DS3, and (c) DS1 and DS3. The interaction parameters for each individual dataset DS1, DS2, and DS3 are obtained by minimizing simultaneously the $\chi^2$ and the length of the solution vector (see Methods). The solid line represents the linear regression between the parameters derived from different datasets. The linear regression correlation coefficient between parameters derived from datasets DS1 and DS2, DS2 and DS3, and DS1 and DS3 are is 0.06, 0.15 and 0.21 respectively.

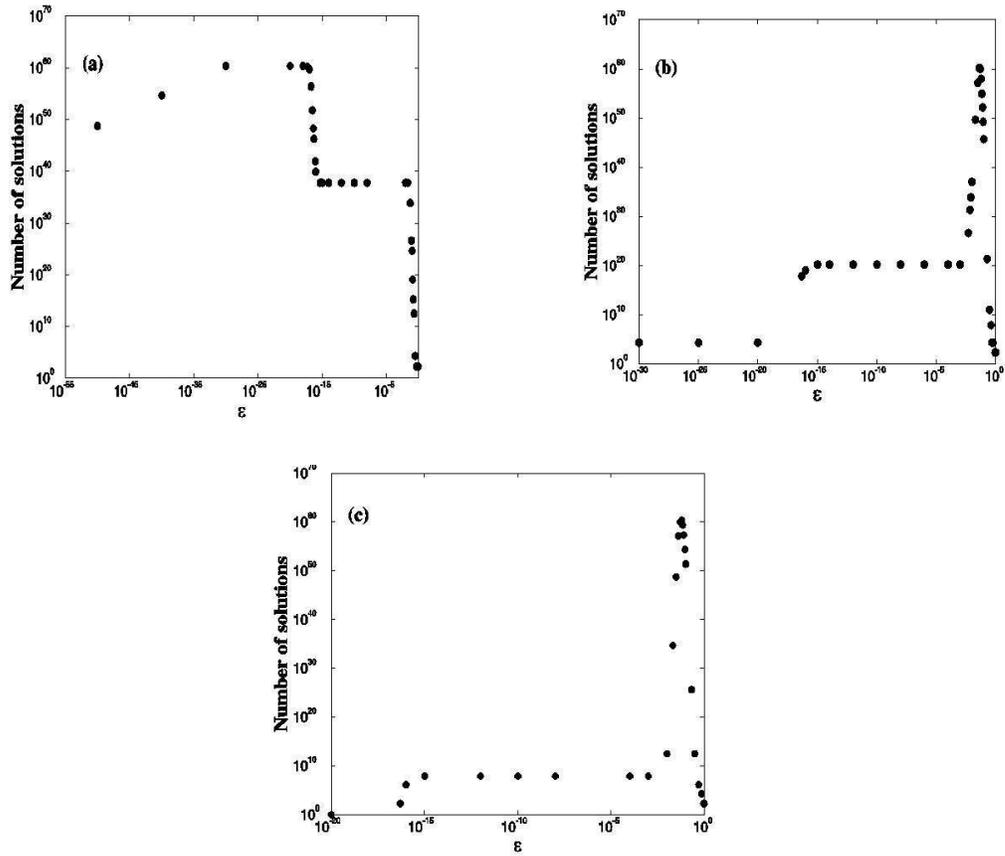

**Figure 6.** The number of solutions as a function of the threshold value ε for datasets (a) DS1, (b) DS2, and (c) DS3. For all datasets, at ε=0, the rank of the contact matrix *A* is equal to the total number of parameters, and we obtain a unique solution. As we increase ε, the rank becomes smaller than the number of parameters, and the number of least square solutions is much greater than 1. When ε=1, the set of linear equations is reduced to one equation, and we obtain $N_p$ solutions.

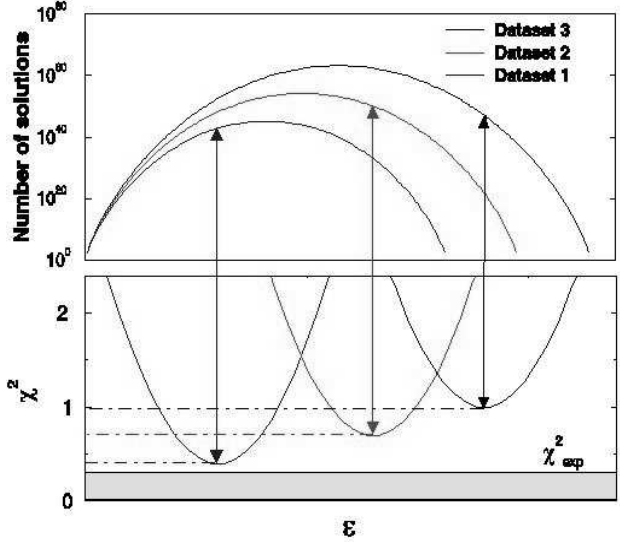

Figure 7. A Cartoon representation of the balance between uniqueness and accuracy of obtained solutions. The parameter ε determines the number of singular values used in SVD and, consequently, the rank of the contact matrix *A* (see Methods). For any set of data, the number of solutions (top panel) is minimal when ε is close to 0 and when ε is close to 1. However, the corresponding $\chi^2$ values, a measure of the accuracy of prediction (bottom panel), is much greater than $\chi^2_{\min}$. The $\chi^2$ values reach a minimum in the range 0<ε<1. We obtain multiple solutions at values of ε corresponding to $\chi^2_{\min}$. In all cases, the obtained $\chi^2_{\min}$ is greater than the error in experimental measurements $\chi^2_{\exp}$. Thus, we argue that a solution that is both unique and accurate ($\chi^2_{\min} \leq \chi^2_{\exp}$) can not be obtained.